\documentclass[12pt]{article}
\usepackage{amsmath,amssymb,amsfonts,amsthm}
\title{Deformed photon-added nonlinear coherent states and their nonclassical properties}

\author{O Safaeian and M K Tavassoly
\\
\footnotesize{Atomic and Molecular Group, Faculty  of Physics,
Yazd University, Yazd, Iran}}
\begin{document}

 \date{\today}

  \maketitle
%===============================================================

 %==============================================================
 \newcommand{\NDD}{\mathcal{N}}
  \newcommand{\HD}{\mathds{H}}
 %%=============================================================

 \begin{abstract}
  In this paper, we will try to present a general formalism for the construction of
  {\it deformed photon-added nonlinear coherent
  states}  (DPANCSs) $\left|\alpha, f, m\right\rangle$, which in special case lead to the well-known photon-added coherent state (PACS)
  $\left|\alpha, m\right\rangle$.
  Some algebraic structures of the introduced DPANCSs are studied and particularly the resolution of the identity, as the most important
  property of generalized coherent states, is investigated. Meanwhile, it will be demonstrated that, the introduced
  states can also be classified in the $f$-deformed coherent states, with a special nonlinearity function.
  Next, we will show that, these states can be produced through a simple theoretical scheme.
  A discussion on the DPANCSs with negative values of $m$, i.e.,
  $\left|\alpha, f, -m\right\rangle$, is then presented.
  Our approach, has the potentiality to be used for the construction of a variety of new
  classes of DPANCSs,
  corresponding to any nonlinear oscillator with known nonlinearity function, as well as arbitrary solvable quantum system with known discrete, nondegenerate spectrum.
  Finally, after applying the formalism to a particular physical system known
  as P\"{o}schl-Teller (P-T) potential and the nonlinear coherent
  states corresponding to a specific nonlinearity function $f(n)=\sqrt n$, some of the nonclassical properties such
  as Mandel parameter, second order correlation function,
  in addition to first and second-order
  squeezing of the corresponding states will be investigated, numerically.
 \end{abstract}
 %===============================================================

 {\bf PACS:} 42.50.-p, 42.50.Dv

%===============================================================
 \vspace{2pc}
 {\bf Keywords:}
   Deformed photon-added nonlinear coherent state, Photon-added coherent state,
   Nonlinear coherent state, Nonclassical state.
%===============================================================

 %===============================================================
  \newpage
 %=====================================================
 \section{Introduction}\label{sec-intro}
 %=====================================================

 As was shown by Agarwal and Tara, "photon-added coherent states" (PACSs) are obtained by iterated actions ($m$ times) of
 bosonic creation operator $a^{\dag}$ on the standard coherent states $\left|\alpha\right\rangle$. The  explicit form of these states
 has been expressed as follows \cite{1}
 \begin{equation}\label{1}
 \left|\alpha, m\right\rangle=\frac{\exp(\frac{-\left|\alpha\right|^2}{2})}{[L_{m}
 (-\left|\alpha\right|^2)m!]^{1/2}}\sum_{n=0}^{\infty}\frac{\alpha^n\sqrt{(n+m)!}}{n!}\left|n+m\right\rangle,
 \end{equation}
 where $m$ is a non-negative integer and $\L_{m}(x)$ is the $m$th-order of Laguerre polynomial. These states
 exhibit nonclassical features like squeezing and sub-Poissonian statistics. Besides, the nonlinear coherent
 state is defined as the solution of the eigenvalue equation
 \begin{equation}\label{2-2}
 A\left|\alpha, f\right\rangle=\alpha\left|\alpha, f\right\rangle,
 \end{equation}
 with the decomposition in the number states space as \cite{2}
 \begin{equation}\label{2}
  \left|\alpha, f\right\rangle=N(\left|\alpha\right|^2)^{-\frac{1}{2}}\sum_{n=0}^{\infty}\frac{\alpha^n}{\sqrt{n!}
   \left[f(n)\right]!}\left|n\right\rangle,
 \end{equation}
  where $n=a^{\dag}a$ is the number operator, $A=af(n)$ is the $f$-deformed annihilation operator, $f(n)$
  is an intensity dependent function,
  $[f(n)]!\dot{=}f(n)f(n-1)...f(1)$ and by convention
  $[f(0)]!\dot{=}1.$ It is shown that the states in (1) may be
  considered as nonlinear coherent states with $f(n,m)=1-m/(1+n)$  \cite{3}.
  Consequently, the eigenvalue equation
  $f(n,m)a\left|\alpha,m\right\rangle=\alpha\left|\alpha,m\right\rangle$
  is established. Using the Stieltjes power-moment
  problem, the over-completeness of PACSs is explicitly shown in
  \cite{4}. Photon-added and photon-subtracted coherent states
  associated with inverse $q$-boson operators introduced in
  \cite{5}. Wave packet dynamics of PACSs has been
  investigated in \cite{6}.
%==================================
  Higher-order squeezing and higher-order sub-Poissonian statistics
  of PACSs have been studied in [7], and squeezing and
  higher-order squeezing of PACSs propagating in a Kerr-like medium
  have been discussed in  [8].
  Dynamical squeezing of
  PACSs were investigated in \cite{9}. Generalized hypergeometric photon-added
  and photon-depleted coherent states introduced
  in \cite{10} and PACSs for exactly solvable Hamiltonian studied in \cite{11}.
  Photon-added Barut-Girardello coherent states
  of the pseudo-harmonic oscillator have been constructed in \cite{12} and recently generation of coherent states
  of photon-added type via pathway of eigenfunctions has been  argued in \cite{penson}.

  On the other side, the experimental scheme for generation of PACSs may be found in the literature.
  Among them, we may refer to \cite{13} and especially to the recent work of Zavatta {\it et al}  in which
  $\left|\alpha,1\right\rangle$ has been produced experimentally
  by using a parametric amplifier consisting of a type I beta-barium borate down-conversion crystal  \cite{14}.

  The goal of the present contribution is to introduce a
  formalism for the construction of {\it
  deformed photon-added nonlinear coherent states} (DPANCSs) by
  iterated actions of "$f$-deformed creation operator" on a
  "nonlinear coherent state". A
  deep insight into the works have been down previously in [1-14], in
  comparison with our work show that, we indeed deformed both
  $a^{\dag}$ (creation operator) and $|\alpha\rangle$ (coherent
  state), respectively
  to $A^{\dag}$ and $\left|\alpha, f\right\rangle.$ Hence, our presentation is essentially different with respect
  to earlier ones, through which we will get new results.

     This paper organizes as follows. After the introduction
     of the explicit form of DPANCSs in the next section, the algebraic structure of the states will be
     investigated in section 3, from which we
     will deduce an appropriate nonlinearity function associated with the introduced states. Section 4 deals with the
     resolution of the identity of the DPANCSs, and then a simple scheme for their generation will be presented  in section 5.
     Then, after applying the proposed approach  to P\"{o}schl-Teller potential (P-T) and the nonlinear coherent
     states corresponding to the nonlinearity function $f(n)=\sqrt n$, as some physical realizations of the formalism,
     the corresponding DPANCSs are introduced and the non-classicality  features of the associated states will be
     numerically investigated, in sections 6. Next, in section 7,  DPANCSs with negative values of $m$ are discussed.
     At last, we conclude the paper in section 8.

%==================================================================================

\section{Introducing the general structure of DPANCSs}

%==================================================================================

  Recall that, the actions of $f$-deformed annihilation and creation operators on the number
  states expressed, respectively by $A\left|n\right\rangle=f(n)\sqrt{n}\left|n-1\right\rangle$ and
  $A^{\dag}\left|n\right\rangle=f(n+1)\sqrt{n+1}\left|n+1\right\rangle$. Following the terminology of
  Solomon in \cite{15}, since we are also working in the deformed quantum optics field and
  the photons annihilate or create by the actions of the relevant $f$-deformed ladder
  operators, the notion of "deformed photon" seems to be reasonable for distinguishing them from usual bosonic counterpart.
  In this section, we want to introduce a new family of coherent states, has been called by us as DPANCS,  using  the definition
 \begin{equation}\label{3}
   \left|\alpha, f, m\right\rangle=N^{m,f}_{\alpha}A^{\dag^m}\left|\alpha,
   f\right\rangle, \qquad  \quad m \in \mathbb{Z}^+,
 \end{equation}
 where $\left|\alpha, f\right\rangle$ is in general any class of nonlinear coherent states introduced in (2)
 and $N^{m,f}_{\alpha}$ is an appropriate normalization constant may be determined.
 It is straightforward to obtain the explicit
  form of DPANCSs in terms of Fock states by standard procedure. The final result read as
 \begin{eqnarray}\label{4}
\left|\alpha, f, m\right\rangle&=&N^{m,f}_{\alpha}\left(\sum^{\infty}_{n=0}\frac{\left|\alpha\right|^{2n}}
{n![f^{2}(n)]!}\right)^{-\frac{1}{2}}\\ \nonumber &\times& \sum^{\infty}_{n=0}\frac{\alpha^n[f(n+m)]
!\sqrt{(n+m)!}}{n![f^{2}(n)]!}\left|n+m\right\rangle,
 \end{eqnarray}
  with the normalization factor
 \begin{equation}\label{5}
N^{m,f}_{\alpha}=\left(\sum^{\infty}_{n=0}\frac{\left|\alpha\right|^{2n}}
{n![f^{2}(n)]!}\right)^{\frac{1}{2}}
\left(\sum^{\infty}_{n=0}\frac{\left|\alpha\right|^{2n}(n+m)![f^{2}(n+m)]!}
{(n!)^2[f^{4}(n)]!}\right)^{-\frac{1}{2}}.
 \end{equation}
 In obtaining (5), we have utilized the relation
 \begin{equation}\label{11}
 A^{\dag^m}=\frac{[f(n)]!}{[f(n-m)]!}a^{\dag^m}.
 \end{equation}
 As a clear fact, notice that, the DPANCSs in (5) reduce to PACSs in (1), when $f(n)=1$.
 It is worth mentioning that the number states $\{ |0\rangle ,|1\rangle , ... ,|m-1\rangle\}$
 are absent from the DPANCSs in (5). This situation is exactly similar to PACSs of Agarwal and Tara \cite{1}.

%===========================================================================
\section{The algebra structure of DPANCSs}
%===========================================================================

  Now, we want to show that, DPANCSs can also be interpreted as $f$-deformed coherent states with a specific
  nonlinearity function.
 This may be done via demonstrating the fact that, the DPANCS may be re-obtained from the eigenvalue equation
 \begin{equation}\label{6}
 f_{d}(n, f, m)a\left|\alpha, f, m\right\rangle=\alpha\left|\alpha, f, m\right\rangle.
 \end{equation}
 Noticing that the non-canonical commutation relation between the $f$-deformed ladder operators read as
 \begin{equation}\label{7}
 [A,A^\dag]=(n+1)f^{2}(n+1)-nf^{2}(n),
 \end{equation}
 accordingly, it is convenient to show that
 \begin{equation}\label{8}
 [A,A^{\dag^m}]=a^{\dag^{m-1}}\frac{[f(n+m-1)]!}{[f(n)]!}[(n+m)f^{2}(n+m)-nf^{2}(n)].
 \end{equation}
 Next, due to the identity
 \begin{equation}\label{9}
 a^{\dag^m}f(n)=f(n-m)a^{\dag^m},
 \end{equation}
 the right-hand side of Eq. (10) can be converted to
 \begin{equation}\label{10}
 [A,A^{\dag^m}]=g(n, m)A^{\dag^{m-1}},
 \end{equation}
 where the relation (7) is used and we have set
 \begin{eqnarray}\label{11-1}
 g\left(n, m\right)\equiv (n+1)f^{2}(n+1)-(n-m+1)f^{2}(n-m+1). \nonumber
 \end{eqnarray}
 Multiplying both sides of equation (2) from the left by $A^{\dag^m}$ yields
 \begin{equation}\label{12}
 A^{\dag^m}A\left|\alpha, f\right\rangle=\alpha A^{\dag^m}\left|\alpha, f\right\rangle.
 \end{equation}
 The commutation relation in (12) helps us to rewrite the latter equation as
 \begin{equation}\label{13}
    \left(AA^{\dag^m}-g(n, m)A^{\dag^{m-1}}\right)\left|\alpha, f\right\rangle=\alpha A^{\dag^m}
    \left|\alpha, f\right\rangle.
 \end{equation}
 At last, making use of the identity $A^{\dag^{-1}}=\frac{1}{(n+1)f(n+1)}a$ \cite{16} leads us
 to the following eigenvalue equation
 \begin{eqnarray}\label{14}
 \left(f(n+1)-\frac{g(n, m)}{(n+1)f(n+1)}\right)a\left|\alpha, f, m\right\rangle=\alpha \left|\alpha, f,
 m\right\rangle.
 \end{eqnarray}
 Comparing Eqs. (15) and (8), gives the form of the nonlinearity function associated to DPANCS
 as follows
 \begin{equation}\label{15}
    f_{d}(n, f, m)=\frac{(n-m+1)f^{2}(n-m+1)}{(n+1)f(n+1)},
 \end{equation}
 where the nonlinearity function $f(.)$ appeared in the right hand side of
 (\ref{15}) is determined by the nonlinearity of the original nonlinear coherent states, $\left|\alpha, f,
 m\right\rangle$ in (\ref{4}).
 So, we have finally succeeded in establishing the DPANCSs as
 $f_d$-deformed coherent states, too. Clearly, setting  $f(n)=1$ in (\ref{15}) one readily obtains
 $f_{d}(n, f, m)=1-\frac{m}{(n+1)}$, which is the nonlinearity function of PACSs \cite{3}.
 %==================================================================

 \section{Resolution of the identity of DPANCSs}

 %==================================================================
 We noticed that, DPANCS in (5) is a superposition of all number states starting with $\left|m\right\rangle$.
 %=========================================
Following the path of  \cite{4, 12}, the unity operator in such a {\it subspace} of the total Hilbert
space, spanned by the basis
$\left\{|n\rangle\right\}_{n=m}^\infty$ has been written as
 \begin{equation}\label{32}
 \hat{I}^{(m)}=\sum^{\infty}_{n=m}\left|n\right\rangle\left\langle n\right|=\sum^{\infty}_{n=0}
 \left|n+m\right\rangle\left\langle n+m\right|.
 \end{equation}
To be precise, the name unity operator for $\hat I^{(m)}$ seems to
be unsuitable and it is more reasonable to be called the
projection operator on the relevant subspace. This operator is
bounded and positive valued with a densely defined inverse \cite
{ali_dense}.

 So, in such a case which we deal with, the (generalized) resolution of the identity takes the form
 \begin{equation}\label{33}
 \frac{1}{\pi}\int\!\!\!\int_{D} d^{2}\alpha \;W\left(\left|\alpha\right|^{2}\right)\left|\alpha, f,
  m\right\rangle\left\langle \alpha, f, m\right|=\hat{I}^{(m)},
 \end{equation}
 where $W\left(\left|\alpha\right|^{2}\right)$ is a positive weight function and $D$ expresses the
 domain of the coherent states centered at the origin of complex plane, both of which may be appropriately determined.
 Generally, $D$ may be entire plane or a finite disk centered at
 the origin, depending on the particular chosen $f(n)$. However, since in the continuation of the paper we will  deal with the
 first type, in what follows we have set infinity in the upper
 bounds of the integrals, i.e., the Stieltjes moment problem has been encountered.
 Here, $\alpha=re^{i\varphi}$ and $d^{2}\alpha \doteq rdrd\varphi.$ By substituting equation
 $\left(5\right)$ into $\left(18\right)$, one  obtains
 \begin{eqnarray}\label{34}
 2\sum^{\infty}_{n=0} &\int^{\infty}_{0}& dr\; r^{2n+1}
 W\left(r^{2}\right)\left(N^{m,f}_{\alpha}\right)^{2}\left(\sum^{\infty}_{n=0}\frac{r^{2n}}
{n![f^{2}(n)]!}\right)^{-1}\\ \nonumber &\times& \frac{(n+m)![f^{2}(n+m)]!}
 {(n!)^2[f^{4}(n)]!}\left|n+m\right\rangle\left\langle n+m\right|=\hat{I}^{(m)},
 \end{eqnarray}
 where we have utilized $\int^{2\pi}_{0} d\varphi e^{i \varphi (n-n^{'})}=2\pi\delta_{nn^{'}}.$
 Considering the following expression for weight function:
 \begin{equation}\label{35} W\left(r^{2}\right)=
 {\left(N^{m,f}_{\alpha}\right)^{-2}}\left(\sum^{\infty}_{n=0}\frac{r^{2n}}
{n![f^{2}(n)]!}\right)r^{2m}\tilde{W}\left(r^{2}\right),
 \end{equation}
 we may rewrite (19) as
 \begin{eqnarray}\label{36}
 2\sum^{\infty}_{n=0} &\int^{\infty}_{0}& dr\; r^{2n+2m+1}\tilde{W}\left(r^{2}\right)\frac{(n+m)!
 [f^{2}(n+m)]!}{(n!)^2[f^{4}(n)]!} \\ \nonumber &\times&\left|n+m\right\rangle\left\langle n+m\right|=\hat{I}^{(m)}.
 \end{eqnarray}
 Obviously, to satisfy this equation, the following moment integral should hold
 \begin{equation}\label{33-33}
 2 \int^{\infty}_{0} dr\; r^{2n+2m+1}\tilde{W}\left(r^{2}\right)=\frac{(n!)^2[f^{4}(n)]!}{(n+m)![f^{2}(n+m)]!}.
 \end{equation}
 Finally, after performing the variable change $r^{2}=x$ and replacing $n+m$ by $k-1$,  we arrive at
 \begin{equation}\label{37}
 \int^{\infty}_{0} x^{k-1}\tilde{W}\left(x\right)dx=\frac{\left[\left(k-m-1\right)!\right]^2
 [f^{4}(k-m-1)]!}{(k-1)![f^{2}(k-1)]!}.
 \end{equation}
 As is clear, prior to investigating this property the explicit form of nonlinearity function, i.e.,
 the particular physical system must be specified.

 %==================================================================

 \section{Generation of the DPANCSs}

 %==================================================================
 In order to produce the DPANCSs in (5) physically, we consider the slab of excited two-level
 atoms through a cavity. Let, the initial state of the atom-field system is expressed by $\left|\Psi(0)\right\rangle=
 \left|\alpha, f\right\rangle\left|e\right\rangle,$ where $\left|e\right\rangle$ is the excited state
 of the atom and $\left|\alpha, f\right\rangle$ is the nonlinear coherent state field. The interaction
 Hamiltonian assumes to have the following configuration
 \begin{eqnarray}\label{26}
 \mathcal{H}&=& \hbar g(\sigma_{+}A+A^\dag\sigma_{-})
 \end{eqnarray}
  where $A$, $A^{\dag}$ are the $f$-deformed ladder operators and $\sigma_{+}$, $\sigma_{-}$ are respectively
  the raising and lowering operators of atomic states.
  In other words, a deeper insight into our proposed Hamiltonian
  in (\ref{26}) shows that we have changed the coupling constant
  $g$ to an alterative coupling $gf(n)$, i.e., our setup works
  with an intensity dependent coupling.
  The initial state $\left|\Psi(0)\right\rangle$
  evolves in time according to
 \begin{equation}\label{255} \left|\Psi(t)\right\rangle=\exp[-i\eta(\sigma_{+}A+A^\dag\sigma_{-})]
 \left|\Psi(0)\right\rangle,
 \end{equation}
 where we have set $\eta\equiv gt$ and $g$ is the coupling constant. For $\eta<<1$
 one has
 \begin{equation}\label{27}
 \left|\Psi(t)\right\rangle\tilde{=}\left(1-i\eta(\sigma_{+}A+A^\dag\sigma_{-})\right)\left|\alpha,
  f\right\rangle\left|e\right\rangle.
 \end{equation}
 Thus, we will have the simple form of the state vector of the whole atom-field system as follows
 \begin{equation}\label{28}
 \left|\Psi(t)\right\rangle=\left|\alpha, f\right\rangle\left|e\right\rangle-i\eta A^\dag\left|\alpha,
  f\right \rangle\left|g\right\rangle.
 \end{equation}
 Therefore, if the atom is detected in the ground state $\left|g\right\rangle$, then the state of
  the field is transferred to $A^\dag\left|\alpha, f\right\rangle$,  which is indeed the DPANCS $\left|\alpha,
   f, 1\right\rangle.$
   Hence,  we could, in principle, produce the state $\left|\alpha, f, 1\right\rangle$.
   Generalizing the above procedure, one can easily produce, in principle, DPANCSs with arbitrary values of $m$, by using the Hamiltonian
 \begin{equation}\label{29}
    \mathcal{H} _{m}=\hbar g(\sigma_{+}A^{m}+A^{\dag^m}\sigma_{-}).
 \end{equation}
 Clearly, the state $|\alpha ,f,m\rangle $ can be produced using an appropriate $m$-photon medium.

 %===========================================================================================

 \section{Physical properties of the DPANCSs}\label{weight}

 %===========================================================================================
 In this section we briefly explain some of
the ordinarily helpful criteria in the relevant literature, which will be used for
investigating the non-classicality exhibition of our introduced
states. Along this purpose, we refer to the sub-Poissonian
statistics, antibunching phenomenon, quadrature squeezing and
finally amplitude squared squeezing. A common feature of all of
the above criteria is that the corresponding Glauber Sudarshan
P-function of a non-classical state is not positive definite. But,
we would like to imply that finding this function is usually a
hard task to do. Altogether, each of the above effects, which will
be considered in the paper, is in fact, sufficient for a quantum state to
belong to non-classical states.
%============================================================================

 \subsection{Non-classicality criteria}\label{6-1}

 %============================================================================

 Now, we are ready to introduce some of the non-classicality signs which are widely used in the literature.
 They will help us to investigate the non-classicality features of the introduced states in (\ref{4}),  corresponding to any chosen physical system.

 \begin{itemize}

 \item{}
  Photon-counting statistics of the states is investigated by evaluating Mandel parameter has been defined as \cite{21}
  \begin{equation}\label{19}
  Q=\frac{\left\langle n^{2}\right\rangle-\left\langle n\right\rangle^{2}}{\left\langle n\right\rangle}-1.
  \end{equation}
 The states for which $Q=0$, $Q<0$ and $Q>0,$ respectively corresponds to Poissonian (standard coherent states),
  sub-Poissonian (non-classical states) and super-Poissonian (classical states) statistics.

\item{}
   Although there are quantum states in which supper-/sub-Poissonian statistical
   behavior is appeared with bunching/antibunching effect, but this is not absolutely true.
   To investigate bunching or antibunching effects, second-order
   correlation function is widely used,  which is defined as follows \cite{Glauber2}:
 \begin{equation}\label{g2(0)2}
   g^{2}(0)= \frac{\langle{a^\dag}^2\;a^2\rangle}{\langle{a^\dag}\;a\rangle^2}.
 \end{equation}
   Depending on the specific nonlinearity function $f(n)$, has been chosen for the construction
   of coherent states,  $g^{2}(0)>1$ ($g^{2}(0)<1$)
   indicates  to bunching (antibunching) effect. The case $g^{2}(0)=1$
   corresponds particularly to the canonical coherent states.

 \item{}
  In order to examine the quantum fluctuations of quadratures of the field, we consider the hermitian operators
  $x=(a+a^\dag)/\sqrt{2}$ and $p=(a-a^\dag)/i\sqrt{2}$
  with commutation relation $[x,p]=i$.  With the help of common definitions of variances of position and
  momentum, the following parameters  may
  be defined: $s_{x} = \frac{(\Delta x)^{2}-0.5}{0.5}$
  and $s_{p}=\frac{(\Delta p)^{2}-0.5}{0.5}$, respectively for quadrature squeezing in $x$ and $p$.
  These squeezing parameters can be re-written as follow:
 \begin{equation}\label{24}
 s_{x}=2\left\langle a^\dag a\right\rangle+\left\langle a^{2}\right\rangle+\left\langle
  a^{\dag^2}\right\rangle-\left\langle a\right\rangle^{2}-\left\langle a^\dag\right\rangle^{2}-2\left\langle
   a\right\rangle\left\langle a^\dag\right\rangle
 \end{equation}
 and similarly for $p$ as
 \begin{equation}\label{25}
 s_{p}=2\left\langle a^\dag a\right\rangle-\left\langle a^{2}\right\rangle-\left\langle
 a^{\dag^2}\right\rangle+\left\langle a\right\rangle^{2}+\left\langle a^\dag\right\rangle^{2}-2\left
 \langle a\right\rangle\left\langle a^\dag\right\rangle.
 \end{equation}
 A state is squeezed in $x$ or $p$ if it satisfies the
 inequalities $-1\leq s_{x}<0$ or $-1\leq s_{p}<0$, respectively.

 \item{}
 Amplitude-squared squeezing \cite{22} is defined in terms of hermitian operators
 $X=\left(a^{2}+a^{\dag ^{2}}\right)/2$ and $P=\left(a^{2}-a^{\dag ^{2}}\right)/2i.$
 Their commutation relation calculated as $[X,P]=i\left(2n+1\right)$.
 The squeezing condition in $X$ or $P$
 are respectively given by $-1\leq S_{X}<0$ or $-1\leq S_{P}<0$, where:
 \begin{eqnarray}
 S_{X}=\frac{(\Delta X)^{2}-\langle n+\frac{1}{2}\rangle}{\langle n+\frac{1}{2}\rangle},\\ \nonumber
 S_{P}=\frac{(\Delta P)^{2}-\langle n+\frac{1}{2}\rangle}{\langle n+\frac{1}{2}\rangle}.
 \end{eqnarray}
The numerator of above parameters can be re-written as follows:
 \begin{eqnarray}\label{45}
 (\Delta X)^{2}-\langle n+\frac{1}{2}\rangle &=&\frac{1}{4}\left(\left\langle a^{\dag^4}\right\rangle+\left\langle a^{4}\right\rangle+2\left\langle
  a^{\dag^2}a^{2}\right\rangle\right)\\ \nonumber
  &-&\frac{1}{4}\left(\left\langle a^{\dag^2}\right\rangle+\left\langle
  a^{2}\right\rangle\right)^{2},
 \end{eqnarray}
 \begin{eqnarray}\label{46}
 (\Delta P)^{2}-\langle n+\frac{1}{2}\rangle &=&\frac{1}{4}\left(-\left\langle a^{\dag^4}\right\rangle-\left\langle a^{4}\right\rangle+2\left
 \langle a^{\dag^2}a^{2}\right\rangle\right) \\ \nonumber &+&\frac{1}{4}\left(\left\langle a^{\dag^2}
 \right\rangle-\left\langle a^{2}\right\rangle\right)^{2}.
 \end{eqnarray}
\end{itemize}

 All of the necessary expectation values for computing $Q$, $g^{2}(0)$, $s_x$,
 $s_p$, $S_X$ and $S_P$, corresponding to  DPANCSs
 with arbitrary nonlinearity function $f(n)$ have been presented in the Appendix A.
%=================================================================================================

 \subsection{Physical properties of the DPANCSs associated with P\"{o}schl-Teller (P-T) potential}
 %============================================================================
 In this subsection, we want to apply the presented mathematical-physics structure of DPANCSs in section (2) to a well-known physical system,
 i.e., P-T potential, which has its specific importance in atomic
 and molecular physics (see \cite{17} and references therein).
 This system possesses the following non-degenerate spectrum $e_{n}=n(n+\nu),\;\; \nu > 2$.
 The special case $\nu=2$ characterizes the one dimensional square potential well.
 The nonlinearity function corresponding to this system according to the proposed formalism in
 \cite{18,19} may be easily obtained as
 \begin{equation}\label{pt1}
 f\left(n\right)=\sqrt{n+\nu}.
 \end{equation}
  Inserting (36) in (5), one can easily create the explicit form of DPANCSs associated to P-T potential.
  We continue our study by discussing some
  of the quantum statistical properties and non-classicality features of the DPANCSs associated with
  the mentioned system. This investigation seems to be necessary, due to the fact that even though the nonlinear coherent states
  mostly possess less or more of the non-classicality signs,
  but there exists nonlinear coherent states which do not show neither of the usual non-classicality criteria \cite{exp}.
  Altogether, before paying attention to this
  subject, we would like to establish the resolution of the identity requirement for
  the introduced states.
 %============================================================================
 \subsubsection{Resolution of the identity for the DPANCSs of P-T potential:}
 %============================================================================

 Due to the central importance of the resolution of the
 identity for any class of coherent states, we examine this property according to (23) by using the nonlinearity
 function of P-T potential, i.e.,
 \begin{equation}\label{37}
 \int^{\infty}_{0} x^{k-1}\tilde{W}\left(x\right)dx=
 \frac{\left(m+\nu\right)![\Gamma\left(k-m\right)]^{2}[\Gamma\left(k-m+\nu\right)]^{2}}
 {(\nu!)^{2}\Gamma\left(k+\nu\right)\Gamma\left(k\right)}.
 \end{equation}
 With the help of definition of Meijer's $G$-function, it follows that \cite{20}
 \begin{eqnarray}\label{38}
 \int^{\infty}_{0}dx\;x^{k-1} &G^{m,n}_{p,q}&\!\left(\begin{array}{ll}
               \hspace{-2mm} \beta x &\hspace{-2mm}\left| \begin{array}{l}
                             a_{1},\;...,\;a_{n},\;a_{n+1},\;...,\;a_{p} \\
                             b_{1},\;...,\;b_{m},\;b_{m+1},\;...,\;b_{q}
                           \end{array}\right.
              \end{array}\hspace{-2mm}\right)
 \\ \nonumber &=&\frac{1}{\beta^{k}}\frac{\prod^{m}_{j=1}\Gamma\left(b_{j}+k\right)\prod^{n}_{j=1}
 \Gamma\left(1-a_{j}-k\right)}{\prod^{q}_{j=m+1}\Gamma\left(1-b_{j}-k\right)
 \prod^{p}_{j=n+1}\Gamma\left(a_{j}+k\right)}.
 \end{eqnarray}
 Comparing equations $\left(37\right)$ and $\left(38\right)$, it is easy to obtain
 \begin{equation}\label{39}
    \tilde{W}\left(x\right)= \frac{\left(m+\nu\right)!}{\left(\nu!\right)^{2}}\hspace{1mm}
    G^{4,0}_{2,4}\!\left(\begin{array}{ll}
               \hspace{-1mm}  x &\hspace{-1mm}\left| \begin{array}{l}
                             0,\;\nu \\
                             -m,\;-m,\;\nu-m,\;\nu-m
                           \end{array}\right.
              \end{array}\hspace{-1mm}\right).
 \end{equation}
 Therefore, via using the above results in (20) and after setting $|\alpha|^2 = x$, the weight function finally takes the form
 \begin{eqnarray}\label{40}
    W\left(x\right)=G^{1,2}_{2,4}\!\left(\begin{array}{ll}
               \hspace{-1mm}  -x &\hspace{-1mm}\left| \begin{array}{l}
                             -m,\;-\nu-m \\
                             0,\;0,\;-\nu,\;-\nu
                           \end{array}\right.
              \end{array}\hspace{-1mm}\right)
              G^{4,0}_{2,4}\!\left(\begin{array}{ll}
               \hspace{-1mm}  x &\hspace{-1mm}\left| \begin{array}{l}
                             m,\;\nu+m \\
                             0,\;0,\;\nu,\;\nu
                           \end{array}\right.
              \end{array}\hspace{-1mm}\right),
 \end{eqnarray}
 where we have utilized the relation \cite{12}
 \begin{eqnarray}\label{40-2}
 x^{s}G^{m,n}_{p,q}\!\left(\begin{array}{ll}
               \hspace{-1mm}  x &\hspace{-1mm}\left| \begin{array}{l}
                             \left(a_{p}\right) \\
                             \left(b_{q}\right)
                           \end{array}\right.
              \end{array}\hspace{-1mm}\right)=G^{m,n}_{p,q}\!\left(\begin{array}{ll}
               \hspace{-1mm}  x &\hspace{-1mm}\left| \begin{array}{l}
                             \left(a_{p}+s\right) \\
                             \left(b_{q}+s\right)
                           \end{array}\right.
              \end{array}\hspace{-1mm}\right)
 \end{eqnarray}
 and
 \begin{eqnarray}\label{40-1}
 N^{m,f}_{\alpha}&=&\left[\nu!\; x^{-\nu/2}
 I_{\nu}\left(2 \sqrt x)\right)\right]^{\frac{1}{2}}\\
 \nonumber &\times&\left[\frac{\left(\nu!\right)^{2}}{\left(m+\nu\right)!}G^{1,2}_{2,4}\!\left(\begin{array}{ll}
               \hspace{-1mm}  -x &\hspace{-1mm}\left| \begin{array}{l}
                             -m,\;-\nu-m \\
                             0,\;0,\;-\nu,\;-\nu
                           \end{array}\right.
              \end{array}\hspace{-1mm}\right)\right]^{-\frac{1}{2}},
 \end{eqnarray}
 which is indeed the closed form of normalization factor for the DPANCSs corresponding to P-T
 potential, and $I_{\nu}(x)$ is the modified Bessel function of the first kind.

 %================================================================

 \subsubsection{Numerical results of the DPANCSs associated with P-T potential:}

 %===============================================================
  Using the mentioned criteria in subsection \ref{6-1}, we will argue and investigate
  the non-classicality of DPANCSs associated with P-T potential. This can be considered as a
  physical realization of our proposed formalism.
  Firstly, figure 1 displays weight function versus $x$ for different values of $m$ and fixed value of $\nu=3.$
  These results signify the positivity of $W\left(x\right)$. It is seen that $W\left(x\right)$
  has a singularity at $x=0$ but it tends to zero for $x\rightarrow \infty$. In figure 2, $W\left(x\right)$
  has been plotted versus $x$ for fixed $m=1$ and various values of $\nu$. The general behavior of weight function in this
  case is the same as figure 1. We continue with investigating the non-classicality of the introduced states. For this purpose, by
  using the required mean values (see Appendix A), we have plotted Mandel parameter,
  second-order correlation function, quadrature squeezing and amplitude-squared squeezing for the DPANCSs of P-T potential versus
  real $\alpha$, for various values of $m$ and fixed $\nu=3$.
  As is shown in figure 3, Mandel
  parameter always is negative and so the sub-Poissonian behavior is visible. It is clearly
  seen that this parameter for the DPANCSs of P-T potential are more negative than
  PACSs ($f(n)=1$). So, our deformation increases the depth of the  non-classicality of these states.
  Besides, increasing $m$ results in
  increasing the non-classicality of DPANCSs in analogously to the results of the PACSs. Meanwhile,
  for large values of $\alpha$ in both of PACSs and DPANCSs, the Q parameters coincide with each other for different chosen  values of $m$.
  It indeed tends to a finite negative value for large $\alpha.$
  %=====================================================
  According to figure 4, it is visible that $g^{2}(0)<1$ for small values of $\alpha$ and so antibunching effect occurs.
  This observation illustrates that sub-Poissonian statistics and antibunching effect
  occur simultaneously in this finite range, as one may compare figures 3 and 4.
  But, our further calculations for larger $\alpha$ (and certainly the same fixed parameters) show that
  while Mandel parameter tends to a finite negative value ($\approx -0.5$),
  the correlation function tends to $\approx  1$, corresponds to  correlation function of canonical coherent state. So,
  while the states in hand have  sub-Poissonian statistics, they  do not  show
  antibunching effect for large $\alpha$.
  It is noticeable that, in this case,
  comparing the two distinct non-classicality  criteria,
  the Mandel parameter is more sensitive than the second-order correlation function.
  %======================================================
  Squeezing parameters have been plotted in figure 5. We conclude from our
  numerical results which  presented in figure 5-a that, the DPANCSs are squeezed in $x$-quadrature, in a wide region
  of $\alpha$, with no squeezing in $p$-quadrature (see figure 5-b). It is evident that for large values of $\alpha$,
  $s_{x}$ and $s_{p}$ respectively tends to $-0.5$ and $1$.
  From figure 5-c it is visible that in some regions of space  $S_{X}$ gets negative values,
  i.e., amplitude-squared squeezing in $X$ appears.  But, as it is
  observed from figure 5-d, $S_{P}$
  is always positive, i.e., no amplitude-squared squeezing in $P$ may be seen.
  Our further calculations confirm that with increasing the values of $\alpha$, $S_{X}$ and $S_{P}$
  respectively tend to $\approx -0.5$  and $\approx 1$.
  Obviously, all of the limiting quantities are correct for the mentioned fixed parameters.

 %===========================================================================
\subsection{Numerical results of the DPANCSs for a nonlinear coherent state with  $f(n)=\sqrt n$}

As second example, we work with the original nonlinear coherent
states corresponding to the nonlinearity function $f(n)=\sqrt n$.
The physical interest in the nonlinear coherent states constructed
by this function comes out from the fact that it, indeed, rises in
a natural way in Hamiltonians illustrating the interaction with
intensity-dependent coupling between a two-level atom and an
electromagnetic radiation field \cite{sing, sukumar}. Considering
this nonlinearity function, the weight function may be
straightforwardly obtained as follows:
\begin{eqnarray}\label{40}
    W\left(x\right)=(m!)^{2} \;_2F_3(1+m,1+m;1,1,1;x)
              G^{4,0}_{2,4}\!\left(\begin{array}{ll}
               \hspace{-1mm}  x &\hspace{-1mm}\left| \begin{array}{l}
                             m,\;m \\
                             0,\;0,\;0,\;0
                           \end{array}\right.
              \end{array}\hspace{-1mm}\right),
 \end{eqnarray}
  where ${}_{p}F_q(a;b;x)$ is the generalized hypergeometric function.
  Figure 6 displays the weight function versus $x$ for different values of $m$.
  The positivity of $W\left(x\right)$ is revealed which confirms that the obtained DPANCS are in fact of coherent states type, in its exact meaning.

  Our aim is producing the DPANCSs associated with this particular system and investigate
  their physical properties. Inserting the function $f(n)=\sqrt n$ in (5),
  one can easily create the explicit form of the associated DPANCSs.
  To proceed further, one needs to use the relations which are presented in Appendix A, setting $f(n)=\sqrt n$.
  For this purpose,  the normalization factor of the related DPANCSs is  required, which may be determined as:
 \begin{eqnarray}
   N_{\alpha}^{m,f}=\sqrt{\frac{I_{0}(2\sqrt{x})}{m!}}\;[\;_2F_3(1+m,1+m;1,1,1;x)\;]^{-1/2},
 \end{eqnarray}
   where  $I_{0}(x)$ is the modified Bessel function of the first kind and ${}_{p}F_q(a;b;x)$ is the generalized hypergeometric function.
   Henceforth, we are now ready to continue with investigating the non-classicality of the associated states.
   We have plotted Mandel parameter, second-order correlation function,
   quadrature squeezing and amplitude-squared squeezing for the corresponding DPANCSs versus
   real $\alpha$, for various values of $m$.
  Mandel  parameter, has been shown in figure 7, is always negative and so the sub-Poissonian behavior is visible. It is clearly
  seen that this parameter for the DPANCSs of the chosen nonlinearity function is more negative than
  PACSs ($f(n)=1$). Therefore, our new deformation also increases the depth of the  non-classicality.
  Besides, increasing $m$ results in
  increasing the non-classicality of DPANCSs in analogously to the numerical results of PACSs and DPANCSs for P-T potential.
  With increasing $\alpha$ in both of PACSs and DPANCSs, the corresponding $Q$ parameters coincide with each other for different chosen  values of $m$.
  Interestingly, it is worth noticing that, while in the case of PACSs for large  $\alpha$,
  $Q$ tends to zero (non-classicality disappears), this is not so for DPANCSs,
  again showing the strong non-classicality behavior of the introduced states.
  %=====================================================
  According to figure 8, it is visible that $g^{2}(0)<1$ for enough small values of $\alpha$ and so antibunching effect will be appeared.
  This observation illustrates that sub-Poissonian statistics and antibunching effect
  occur simultaneously in this finite range, as one may compare figures 7 and 8.
  But, our further calculations for larger $\alpha$ show that
  while Mandel parameter tends to a finite negative value ($\approx -0.6$),
  the correlation function tends to $\approx  1$, corresponds to correlation function of canonical coherent state. So,
  the presented states have sub-Poissonian statistics with no
  antibunching effect for large $\alpha$.
  Therefore, we may conclude that  comparing the above two non-classicality  criteria,
  Mandel parameter is more sensitive than the second-order correlation function.
  %======================================================
  Squeezing parameters have been plotted in figure 9. It is obvious from our
  numerical results presented in figure 9-a that the corresponding DPANCSs are squeezed in $x$-quadrature, in a wide region
  of $\alpha \geq 1.75$, while no squeezing is seen in $p$-quadrature (see figure 9-b).  It is also evident that for large values of $\alpha$,
  $s_{x}$ and $s_{p}$ respectively tend to $-0.5$ and $1$, for those chosen values of $m$.
  %======================================================
  From figure 9-c it is visible that in some regions of space, especially large values of  $\alpha$, $S_{X}$ gets negative values,
  i.e., amplitude-squared squeezing in $X$ appears.  But, as it is
  observed from figure 9-d, $S_{P}$
  is always positive, i.e., no amplitude-squared squeezing in $P$ may be seen.
  Our further calculations show that, at leat for the chosen parameters, with increasing the values of $\alpha$, $S_{X}$ and $S_{P}$
  respectively tend to $\approx -0.5$  and $\approx 1$.

 %===========================================================================

 \section{A discussion on DPANCSs with negative $m$}

 %===========================================================================
 The form of $f_{d}(n, f, m)$ in (16) suggests that, a nonlinearity
 function can be constructed also for negative integer values of $m$. The corresponding coherent states
 associated with this nonlinearity function, denoted by us as
 $\left|\alpha, f, -m\right\rangle$, will be called as "DPANCSs with negative $m$".
  In order to construct these states, consider the following eigenvalue equation
 \begin{equation}\label{16}
 \frac{(n+m+1)f^{2}(n+m+1)}{(n+1)f(n+1)}a\left|\alpha, f, -m\right\rangle=\alpha \left|\alpha, f, -m\right\rangle,
 \end{equation}
 which is obtained simply, by replacing $m$ with $-m$ in (15) together with (16).
 Following the usual procedure, i.e., by expanding $\left|\alpha,
 f, -m\right\rangle$ in terms of the number states and finding the
 expansion coefficients, one straightforwardly arrives at
 \begin{eqnarray}\label{dpaneg}
 \left|\alpha, f, -m\right\rangle= N^{{-m},f}_{\alpha} \sum^{\infty}_{n=0} \frac{\alpha^{n}m!\sqrt{n!}[f(n)]![f^{2}(m)]!}{(n+m)![f^{2}(n+m)]!}
 \left|n\right\rangle.
 \end{eqnarray}
  The constant $N^{{-m},f}_{\alpha}$ is determined by normalization condition as
 \begin{equation}\label{18}
  N^{-m,f}_{\alpha}=\left(\sum_{n=0}^\infty\frac{\left|\alpha\right|^{2n}n!(m!)^{2}[f^{2}(n)]![f^{4}(m)]!}
  {[(n+m)!]^{2}[f^{4}(n+m)]!}\right)^{-\frac{1}{2}}.
 \end{equation}
 Unlike the DPANCSs in (5), the states $\left|\alpha, f, -m\right\rangle$ contain a superposition of all
 Fock states starting with the vacuum state  $\left|0\right\rangle.$ In the limit
 $\alpha\rightarrow 0$,
 the state $\left|\alpha, f, -m\right\rangle$ reduces to the vacuum state, but in the same limit,
 irrespective of the value of $m$, DPANCS reduces to the number state $\left|m\right\rangle$.
 Also, in the limit $m\rightarrow0$, both of the states $\left|\alpha, f, \pm
 m\right\rangle$,  recover trivially the original nonlinear coherent state  $\left|\alpha, f\right\rangle.$
 It is worth to add the point that the states
 $|\alpha, -m\rangle$ which previously argued in \cite{3}, may be reobtained by setting $f(n)=1$
 in $|\alpha, f, -m\rangle$ in (\ref{dpaneg}).

 %============================================================================
  The procedure which we followed in section (\ref{weight}) for investigating the resolution of the identity and
  obtaining the appropriate weight function of
  DPANCSs, can be used for the DPANCSs with negative $m$ in (\ref{dpaneg}).
  Notice that for these states the well-defined unity operator is as usual
 \begin{equation}\label{32}
 \hat{I}^{(-m)}=\hat{I} = \sum^{\infty}_{n=0}\left|n\right\rangle\left\langle n\right|.
 \end{equation}
 So, in such a case which we deal with, the resolution of the identity requirement takes the form
 \begin{equation}\label{33}
 \frac{1}{\pi}\int\!\!\!\int d^{2}\alpha \;W^{(-m)}\left(\left|\alpha\right|^{2}\right)\left|\alpha, f,
  -m\right\rangle\left\langle \alpha, f, -m\right|=\hat{I},
 \end{equation}
 where $W^{(-m)}\left(\left|\alpha \right|^{2}\right)$ denotes the non-negative weight function should be determined.
 In this way, one straightforwardly gets
 \begin{equation}\label{wbn1}
 \int ^{\infty}_{0}dx\;x^{n}\tilde{W}^{(-m)}(x)=\frac{((n+m)!)^{2}[f^{4}(n+m)]!}{n![f^{2}(n)]!}
 \end{equation}
 where
 \begin{equation}\label{wbn2}
  \tilde{W}^{(-m)}(x)=(N^{-m,f}_{\alpha})^{2}x^{-m}(m!)^{2}[f^{4}(m)]!W^{(-m)}\left(x\right).
 \end{equation}
  Investigating the case for particular physical system, i.e., P-T potential, we finally
  arrive at
 \begin{eqnarray}\label{40}
    W^{(-m)}\left(x\right)&=&\frac{(\nu!)^{2}}{[\left(m+\nu\right)!]^{4}\left(N^{-m,f}_{\alpha}\right)^{2}
     (m!)^{2}} \\ \nonumber &\times& G^{4,0}_{2,4}\!\left(\begin{array}{ll}
      \hspace{-1mm}  x &\hspace{-1mm}\left| \begin{array}{l}
      0,\;\nu \\
       m,\;m,\;\nu+m,\;\nu+m
   \end{array}\right.
  \end{array}\hspace{-1mm}\right),
 \end{eqnarray}
  where $N^{-m,f}_{\alpha}$ has been introduced in (\ref{18}).
  The DPANCSs with negative $m$ may be called "coherent states" (according to the Klauder definition),
  if the weight function $W^{(-m)}\left(x\right)$ will be positive in all space.
  To check this requirement we have plotted $W^{(-m)}(x)$ in figure 10  versus $x$ for fixed value $\nu =3$ and different values of $m$.
  As is shown, unfortunately $W^{(-m)}\left(x\right)$ in some region of space gets negative values.
  So, our results in figure 10 indicate that the DPANCSs with negative $m$, associated to P-T potential can not be known as coherent state.\\
  Upon the latter results, we motivated to check the above
  procedure for the PACSs with negative $m$, as  introduced in
  \cite{3} (the states which may be reproduced by setting $f(n)=1$
  in (\ref{dpaneg})).
  Unfortunately, the same conclusion  has been
  obtained, i.e., $W^{(-m)}_{PACS}(x)$ will get negative values in some
  regions of space (see figure 11). We have also examined our conclusion for the DPANCSs with negative $m$
  associated with other nonlinearity functions. For this purpose we worked with $f(n)=1/\sqrt n$
  (harmonious states \cite{sudarshan}), $f(n)=1/\sqrt{n+2\kappa-1}$ (Barut-Girardello coherent states
  of $SU(1, 1)$ group \cite{Ali}) and $f(n)=\sqrt n$. In all the latter cases we obtained the same result,
  i.e., the positivity of the weight function and so the overcompleteness relation do not justify.
  We should mention that we investigated the uniqueness of the solution of the moment integrals by
  examining the Carleman criterion \cite{carleman}.
  \\We end this section with mentioning another example for the latter result may be found in the
  literature, where the following state has been
  introduced by Klauder {\it et al}  \cite{23}:
 \begin{eqnarray}
 |\alpha\rangle =[\;_2F_2(1,1;2,2;|\alpha|^{2})\;]^{-1/2}\sum_{n=0}^{\infty}\frac{\alpha ^{n}}{\sqrt{(n+1)^{2}n!}}|n\rangle,
 \end{eqnarray}
 with ${}_{p}F_q(a;b;x)$ as the generalized hypergeometric function.
 As is illustrated there, the weight function for this set of states has been derived as
 $W(x)=x e^{-x}(x-1)$, which is not trivially a non-negative function in all space.
 So, even though this class of states is normalizable and continuous in the label,
 however it is not known as a coherent state, in its exact meaning.

  %============================================================
  \section{Summary and conclusion}
  %============================================================
  In summary, we presented a general formalism for the
  construction of DPANCSs with the explicit form in (\ref{4}),
   which recovers, in special cases, PACSs (setting $f(n)=1$ in (\ref{4})) and canonical coherent states (setting $f(n)=1$ in (\ref{4}),
   together with $m=0$).
  The algebraic structure and the resolution of the identity requirement of the introduced states are also illustrated.
  As in the case of PACSs, we established that the DPANCSs can be specified with a
  nonlinearity function denoted by us as $f_d(n, f, m)$.
  Therefore, the introduced DPANCSs are of the $f$-deformed type, too.
  We briefly argued that, their physical generation is possible.
  Then, after applying the formalism to P-T potential,
  the non-classical properties of the DPANCSs associated with P-T potential are checked
  through evaluating Mandel parameter, second-order correlation function,
  as well as first and second-order squeezing, numerically.
  %======================================================================
  Along the physical realization of the formalism, we also briefly
  applied the same procedures which have been done for the P-T potential
  to a well-known class of nonlinear coherent states with $f(n)=\sqrt
  n$. Generally, we observed much intensity (in depth and domain) of non-classicality signs for the DPANCSs associated
  with such system in comparison with PACSs of \cite{1}.
  %======================================================================
  Then, a discussion on the "DPANCSs with negative $m$" is
  presented. According to our results, it is deduced that these latter states
  do not satisfy the resolution of the identity, appropriately.
  Indeed, the function which satisfies the related moment integral uniquely determined,
  altogether the positivity of the obtained function did not
  confirmed. Also, we further investigated the existence and positivity of the weight function for the PACSs
  with negative values of $m$ have been introduced in \cite{3}. Unfortunately, we have found the same conclusion.
  So, recalling the minimal requirements of any quantum state to be exactly named "coherent state" \cite{23},
  we may conclude that the "PACSs and DPANCSs with negative $m$"
  are not strictly known as coherent states.

  %=====================================================================
  %=====================================================================
  Finally, it is worth  mentioning that even though we have used only  P-T potential and a particular class of nonlinear coherent state
  as some physical realizations of our proposed structure,
  its essential potentiality to be used for any class of nonlinear coherent states with known nonlinearity function,
  in addition to any solvable quantum system with arbitrary discrete spectrum should be clear.
  So, in this way, a vast new family of DPANCSs can, in principle, be constructed.
  Apart from the generalized structure of our
  proposal, it is noticeable that,
  it is a rather different formalism with new outputs in resultant coherent states and their non-classicality aspects,
  in comparison with earlier works [1-12].

 %============================================================
 \begin{flushleft}
 {\bf Acknowledgments}\\
 \end{flushleft}
 One of the  authors (O S) wishes to thank  M J Faghihi for useful
 discussions. Also,  the authors are grateful from Professor K A
 Penson for his valuable hints on part of the numerical calculations of the
 paper. At last, we are thankful from the referees for their useful comments which improved the quality of the paper.

 %============================================================
 %============================================================
 %\appendix
 %============================================================
 \section{Appendix}
 %============================================================

 The mean values of the relevant operators over the state
 $\left|\alpha,f,m\right\rangle$,
  required for our numerical calculations, may be easily obtained as follows:
 \begin{eqnarray}\label{21}
 \left\langle a\right\rangle&=&\alpha \left(\tilde{N}^{m,f}_{\alpha}\right)^{2}\sum^{\infty}_{n=0}
 \frac{\left|\alpha\right|^{2n}(n+m)![f^{2}(n+m)]!}{(n!)^{2}[f^{4}(n)]!}\\
  \nonumber &\times&\frac{(n+m+1)f(n+m+1)}{(n+1)f^{2}(n+1)},
 \end{eqnarray}

 \begin{eqnarray}\label{22}
 \left\langle a^{2}\right\rangle &=&\alpha^{2}\left(\tilde{N}^{m,f}_{\alpha}\right)^{2} \sum^{\infty}_
 {n=0}\frac{\left|\alpha\right|^{2n}(n+m)![f^{2}(n+m)]!}{(n!)^{2}[f^{4}(n)]!}\\ \nonumber
 &\times& \frac{(n+m+1)(n+m+2)f(n+m+1)f(n+m+2)}{(n+1)(n+2)f^{2}(n+1)f^{2}(n+2)},
 \end{eqnarray}

 \begin{eqnarray}\label{22-1}
 \left\langle a^{4}\right\rangle &=&\alpha^{4}\left(\tilde{N}^{m,f}_{\alpha}\right)^{2} \\ \nonumber &\times &\sum^{\infty}_
 {n=0}\frac{\left|\alpha\right|^{2n}(n+m+4)![f(n+m+4)]![f(n+m)]!}{n!(n+4)![f^{2}(n)]![f^{2}(n+4)]!},
 \end{eqnarray}

 \begin{equation}\label{23}
 \left\langle a^\dag a\right\rangle=\left(\tilde{N}^{m,f}_{\alpha}\right)^{2}\sum^{\infty}_{n=0}
 \frac{\left|\alpha\right|^{2n}(n+m)![f^{2}(n+m)]!}{(n!)^{2}[f^{4}(n)]!}(n+m),
 \end{equation}

 \begin{eqnarray}\label{23}
 \left\langle a^{\dag^{2}} a^{2}\right\rangle &=&\left(\tilde{N}^{m,f}_{\alpha}\right)^{2} \\ \nonumber &\times & \sum^{\infty
 }_{n=0}\frac{\left|\alpha\right|^{2n}(n+m)![f^{2}(n+m)]!(n+m)(n+m-1)}{(n!)^{2}[f^{4}(n)]!},
 \end{eqnarray}

 \begin{equation}\label{222}
 \left\langle \left(a^\dag a\right)^{2}\right\rangle=\left(\tilde{N}^{m,f}_{\alpha}\right)^{2}\sum^
 {\infty}_{n=0}\frac{\left|\alpha\right|^{2n}(n+m)![f^{2}(n+m)]!}{(n!)^{2}[f^{4}(n)]!}(n+m)^{2},
 \end{equation}
%==============================================================
  where we have set $\tilde{N}^{m,f}_{\alpha}=N^{m,f}_{\alpha}\left(\sum^{\infty}_{n=0}\frac{\left|\alpha\right|^{2n}}
 {n![f^{2}(n)]!}\right)^{-\frac{1}{2}}$ and $N^{m,f}_{\alpha}$
 determined in (\ref{5}). Note that, $\left\langle a^\dag\right\rangle,$
 $\left\langle a^{\dag^2}\right\rangle$ and  $\left\langle a^{\dag^{4}}\right\rangle$ can be obtained
  by taking the complex conjugate of $\left\langle a\right\rangle,$ $\left\langle a^{2}\right\rangle$
  and $\left\langle a^{4}\right\rangle,$ respectively.

%==============================================

 \newpage

%==============================================
 {\bf FIGURE CAPTIONS}
%==============================================

 \vspace {.5 cm}

 {\bf FIG. 1} The plot of $W(x)$ as a function of $x$, with fixed parameter $\nu=3$ and different values of $m$ for
 DAPNCS associated with P-T potential. Continuous line is for $m=1$, dotted line is for $m=5$ and dashed line is for $m=9$.

  \vspace {.5 cm}

 {\bf FIG. 2} The plot of $W(x)$ as a function of $x$, with fixed parameter $m=1$ and different values of $\nu$ for
 DAPNCS associated with P-T potential. Continuous line is for $\nu=7$, dotted line is for $\nu=5$ and dashed line is for $\nu=3$.

  \vspace {.5 cm}

 {\bf FIG. 3} The variation of $Q$ as a function of $\alpha \in \mathbb{R}$, with fixed
              parameter $\nu=3$ and different values of $m$ for
              DPANCS associated with P-T potential. Continuous line is for DPANCSs and $m=2$, dashed line is for DPANCSs and $m=5$, dot-dashed line is for PACSs and $m=2$, and dotted line is for PACSs and $m=5$.

 \vspace {.5 cm}

 {\bf FIG. 4} The variation of $g^{2}(0)$ as a function of $\alpha \in \mathbb{R}$, with fixed
              parameters $\nu=3$ and different values of $m$ for
              DPANCS associated with P-T potential. Continuous line is for $m=1$, dashed line is for $m=2$ and dotted line is for $m=3$.

 \vspace {.5 cm}

 {\bf FIG. 5} Squeezing parameters as a function of $\alpha \in \mathbb{R}$, with
              fixed parameter $\nu=3$ and different values of $m$ for
              DPANCS associated with P-T potential. (a): shows the variation of $s_{x}$, continuous line is for $m=1$, dotted line is for $m=2$ and dashed line is for $m=3$; (b): the same as (a) except that it is plotted for $s_{p}$; (c): shows the variation of $S_{x}$, continuous line is for $m=1$, dotted line is for $m=3$ and dashed line is for $m=5$; (d): the same as (c) except that it is plotted for $S_{p}$.

 \vspace {.5 cm}
 {\bf Fig. 6} The plot of $W(x)$ as a function of $x$ for DAPNCS associated with $f(n)=\sqrt n$.
 Continuous line is for $m=1$, dotted line is for $m=5$ and dashed
line is for $m=9$.

 \vspace {.5 cm}

 {\bf Fig. 7}  The parameter $Q$ as a function of $\alpha \in \mathbb{R}$.
 Continuous line and dashed line  are respectively for $m=2$ and $m=5$ for
 DPANCS associated with $f(n)=\sqrt n$. The dot-dashed line and  dotted
line are respectively for $m=2$ and $m=5$ for PACSs ($f(n)=1$).

 \vspace {.5 cm}

 {\bf Fig. 8} The variation of $g^{2}(0)$ as a function of $\alpha \in \mathbb{R}$ for DPANCS associated with $f(n)=\sqrt n$.
     Continuous line is for $m=1$, dashed line is for $m=2$ and dotted
     line is for $m=3$.

 \vspace {.5 cm}
 {\bf Fig. 9} Squeezing parameters as a function of $\alpha \in \mathbb{R}$ for different values of $m$ for
              DPANCS associated with $f(n)=\sqrt n$. (a): shows the variation of $s_{x}$, continuous line is for $m=1$, dotted line is for $m=2$ and dashed line is for $m=3$; (b): the same as (a) except that it is plotted for $s_{p}$; (c): shows the variation of $S_{x}$, continuous line is for $m=1$, dotted line is for $m=3$ and dashed line is for $m=5$; (d): the same as (c) except that it is plotted for $S_{p}$.

 \vspace {.5 cm}

 {\bf FIG. 10} The plot of $W^{(-m)}(x)$ as a function of $x$, with fixed
               $m$ parameters  and $\nu=3$ for
               DPANCS  with negative $m$  associated with P-T potential. Continuous line is for $m=1$,
               dashed line is for $m=2$ and dotted line is for $m=3$.

 \vspace {.5 cm}

 {\bf FIG. 11}  The plot of $W^{(-m)}_{PACS}(x)$ as a function of $x$, with
                fixed  $m$ parameters for PACS with negative $m$.
                Continuous line is for $m=1$, dotted line is for $m=2$ and dashed line is for $m=3$.

%==============================================================================
 \end{document}